\documentclass{llncs}
\usepackage{lineno}
\usepackage[pagebackref=false,colorlinks=false]{hyperref}
\modulolinenumbers[5]

\usepackage{amsmath,amssymb,amsfonts,thmtools,thm-restate}
\usepackage[normalem]{ulem}
\usepackage{tikz}
\usepackage{todonotes}
\usepackage{thm-restate}

\usepackage[normalem]{ulem}
\usepackage{amsmath}
\usepackage{amssymb}
\usepackage{stmaryrd}
\usepackage{subcaption}
 \usepackage[all]{xy}
\usepackage{bussproofs}
\usepackage{mathtools} 
\usepackage{wasysym}
\usepackage{lscape}
\usepackage{graphicx}
\usepackage{cancel}
\usepackage{proof}

\newcommand{\conn}[1]{\overset{#1}{\Mapsto}}

\newcommand{\obs}[2]{\langle #1\vartriangleright #2\rangle}
\newcommand{\CNA}{\textsf{CNA}}

\newcommand{\nil}{\mathbf{0}}

\newcommand{\defeq}{\triangleq}

\newcommand{\entails}{\models}

\belowcaptionskip\abovecaptionskip

\begin{document}

\title{SOS Rules for Equivalences of Reaction Systems\thanks{
    Research supported
    by MIUR PRIN 201784YSZ5 \emph{ASPRA}, 
    by Univ.\ of Pisa PRA\_2018\_66  \emph{DECLWARE}.
  }
}

\author{ Linda Brodo\inst{1} \and Roberto Bruni\inst{2} \and Moreno 
Falaschi\inst{3}}
\institute{Dipartimento di Scienze economiche e aziendali,
          Universit\`a di Sassari, Italy\\
	   \email{brodo@uniss.it.} 
	   \and
	   Dipartimento di Informatica,
	   Universit\`{a} di Pisa, Italy\\
	   \email{bruni@di.unipi.it.} 
	   \and
           Dipartimento di Ingegneria dell'Informazione e Scienze Matematiche\\
           Universit\`a di Siena, Italy
           \\
           \email{moreno.falaschi@unisi.it.}
	   }
\maketitle

\begin{abstract}

Reaction Systems (RSs) are a successful computational framework inspired by biological systems.
A RS pairs a set of entities with a set of reactions over them.
Entities can be used to enable or inhibit each reaction, and are produced by reactions.
Entities can also be provided by an external context.
RS semantics is defined in terms of an (unlabelled) rewrite system:
given the current set of entities, a rewrite step consists of the application of all and only the enabled reactions.
In this paper we define, for the first time, a labelled transition system for RSs in the structural operational semantics (SOS) style.
This is achieved by distilling a signature whose operators directly correspond to the ingredients of RSs and by defining some simple SOS inference rules for any such operator to define the behaviour of the RS in a compositional way.
The rich information recorded in the labels allows us to define an assertion language to tailor behavioural equivalences on some specific properties or entities.
The SOS approach is suited to drive additional enhancements of RSs along features such as quantitative measurements of entities and communication between RSs.
The SOS rules have been also exploited to design a prototype implementation in logic programming. 
\end{abstract}

\begin{keywords}
 SOS rules, 
 Reaction Systems,
 assertions,
 logic programming 
\end{keywords}


\section{Introduction}
Labelled Transition Systems (LTSs) are a powerful structure to model the behaviour
 of interacting processes.
An LTS can be conveniently defined following the Structural Operational Semantics (SOS) approach~\cite{Plotkin81,DBLP:journals/jlp/Plotkin04a}.
Given a  signature, an SOS system assigns some inference rules to each operator of the language: the conclusion of each rule is the transition of a composite term, which is  determined by those of its constituents (appearing as premises of the rule).
The SOS approach has been particularly successful in the area of process algebras~\cite{Milner80,DBLP:conf/ifip2/Plotkin82,DBLP:phd/ethos/Hillston94}.

Reaction Systems (RSs)~\cite{BEMR11} are a computational framework inspired by systems of living cells.
Its constituents are a finite set of entities and a finite set of reactions acting on entities.
A reaction is a triple $(R,I,P)$ where $R$ is the set of reactants (entities whose presences is needed to enable the reaction), $I$ is the set of inhibitors (entities whose absence is needed to enable the reaction) and $P$ is the set of products (entities that are produced if the reaction takes place and that will be made available at the next step).
After their introduction, RSs have shown to be a quite general computational model 
whose application ranges from the modelling of biological phenomena~\cite{ABP14,CMMBM12,Az17,BarbutiGLM16}, 
and molecular chemistry~\cite{OY16} to theoretical foundations of computing.
The semantics of RSs is defined as an unlabelled rewrite system whose states are set of entities (coming from an external context or produced at the previous step). Given the current set of entities, a rewrite step consists of the application of all and only the enabled reactions.
Given a sequence of entities to be provided by the context at each step, the behaviour of an RS is uniquely determined and the corresponding (unlabelled, deterministic) transition system is finite.

Here we will define, for the first time, an LTS semantics for RSs in the SOS style.
First we fix a process signature whose operators pinpoint the basic structure of a RS.
We have operators for entities and reactions.
For contexts we exploit some classic process algebraic operators (action prefix, sum and recursion). This way we can recursively define contexts that possibly exhibit nondeterministic behaviour, as sometimes have already appeared in the literature~\cite{Kleijn2018,BBF19}.
Even though we enrich the expressiveness of contexts, the overall LTS still remains finite.
The SOS approach has several advantages:
1)~compositionality, the behaviour of each composite system is defined in term of the behaviours of its constituents;
2)~each transition label conveys all the activities connected to that rewrite step;
3)~the definition of contexts is better  integrated in the framework;
4)~different kinds of contexts (recursive, nondeterministic) can be considered;
5)~it is now easier to change or extend the concept of RSs by adding new operators;
6)~SOS rules facilitate implementation in a declarative language and the application of standard techniques for defining equivalences between processes.

The transition labels of our LTS are so rich of information that standard notion of behavioural equivalence (like traces or bisimulation) are too fine grain.
For studying RSs, one is often interested in focussing on some entities and disregard others, like exploiting a microscope to enhance certain details and ignore others that fall out of the picture.
To this aim, following the ideas in our previous paper~\cite{TCS:RSLink}, we  propose an assertion language built over the transition labels, and we make the definition of behavioural and logical equivalences parametric w.r.t. such assertions.
This way, it is possible to consider different RSs as equivalent for some purposes or to distinguish them for other purposes.

Then, by following an approach similar to the one first presented in~\cite{TCS:RSLink} we develop suitable behavioural equivalences for RS processes, and show the correspondence between a coinductive definition in terms of bisimilarity and its logical counterpart \emph{\`a la} Hennessy-Milner.

We have developed a prototype implementation in logic programming of our semantic framework 
available online, as we describe in Section~\ref{sec:impl}. 
Our interpreter allows the user to
check automatically on the labels for a given RS the validity of formulas expressed 
in our variant of the Hennessy-Milner logic 
combined with the assertions specified in our language.

\paragraph{Related work.}
The work by Kleijn et al.~\cite{Kleijn2018} presents an LTS for RS over $2^S$ states, where $S$ is the set of entities.
Two labelled transition system versions have been proposed: state-oblivious context controller, and state-aware context controller.  In the first version, the  transition labels only record the entities provided by the context,  and in the second one the transition labels also provide the entities composing the actual state. The last choice allows one to decide which entities the context should provide.
Differently, we give a process algebra-style definition of the RS, where the SOS rules produce informative transition labels, including  
context specification, 
allowing different kinds of analysis.


There are some previous works based on bisimulation applied to models for biological systems. Barbuti et al.~\cite{BMMT08} define a classical setting for bisimulation for two formalisms: the Calculus of Looping Sequences, which is a rewriting system, and the Brane Calculi, which is based on process calculi.
Bisimulation is used to verify properties of the regulation of lactose degradation in
Escherichia coli and the EGF signalling pathway. These calculi allow the authors to model membranes' behaviour.
Cardelli et al.~\cite{CTTV15} present two quantitative behavioral equivalences over species of a 
chemical reaction network with semantics based on ordinary differential equations.
Bisimulation identifies a partition where each equivalence class represents the exact sum of the concentrations of the species belonging to that class.
Bisimulation also relates species that have identical solutions at all time points when starting from the same initial conditions.
Both the mentioned formalisms~\cite{BMMT08,CTTV15} adopt a classical approach to bisimulation. 

In Brodo et al.~\cite{BBF19,TCS:RSLink} we derived similar results to those presented here by encoding RSs into c\CNA, a multi-party process algebra  (a variant of the {\tt link}-calculus~\cite{BBB17,BODEI2020104587}).
In comparison with the encoding of RS in c\CNA, we get here a much simpler computational model, closer to the syntax of RSs, preserving the expressiveness at the level of transition labels.

\paragraph{Structure of the paper.} 
In Section~\ref{sec:rs} we recall the basics of RSs.
The original contribution starts from Section~\ref{sec:LTSforRS}, where: 1)~we introduce the syntax and operational semantics of a novel process algebra for RSs, 2)~we show how to encode RSs as processes,  3)~we state a tight correspondence between the classical semantics of RSs and the operational semantics of their corresponding processes. Section~\ref{sec:biosimulation} shows the correspondence between a coinductive definition in terms of bisimilarity and its logical counterpart \emph{\`a la} Hennessy-Milner. A prototype implementation in logic programming of our semantic framework is briefly described in Section~\ref{sec:impl}. Further extensions of RSs that build on our theory are sketched in Section~\ref{sec:extension}. Some concluding remarks are in Section~\ref{sec:conc}.

\section{Reaction Systems}
\label{sec:rs}

The theory of Reaction Systems (RSs)~\cite{BEMR11} 
was born in the field of Natural Computing 
to model the behaviour of biochemical reactions in living cells.

We use the term \emph{entities} to denote generic molecular substances (e.g., atoms, ions, molecules) that may be present
in the states of a biochemical system.
The main mechanisms that regulate the functioning of a living cell are 
{\em facilitation} and {\em inhibition}. These mechanisms are 
based on the presence and absence of entities and are reflected in the basic definitions of RSs.

\begin{definition}[Reaction]
Let $S$ be a (finite) set of entities. A reaction in $S$ 
is a triple $a = (R,I,P)$, where $R, I, P\subseteq S$ are finite,  
non empty sets  and
$R \cap I = \emptyset$. 
\end{definition}

The sets $R, I, P$ are the sets of
\emph{reactants},
\emph{inhibitors}, and  \emph{products}, respectively. 
All reactants are needed for the reaction to take place.
Any inhibitor blocks the reaction.
Products are the outcome of the reaction.
Since $R$ and $I$ are not empty, all products are produced from at least one reactant and every reaction can be inhibited. 
%
We let  $\mathit{rac}(S)$
be the set of all reactions in $S$.
%

\begin{definition}[Reaction System]
A Reaction System (RS) is a pair ${\cal A} = (S, A)$ 
s.t. $S$ is a finite set, and $A \subseteq \mathit{rac}(S)$ is a finite set of reactions in $S$.
\end{definition}


The theory of RSs is based on 
three
assumptions:
{\bf no permanency}, any entity vanishes unless it is sustained 
by a reaction. In fact, a living cell would die for lack 
of energy, without chemical reactions;
{\bf no counting}, the basic model of RSs is very abstract and 
qualitative, i.e. the quantity of entities that are present 
in a cell is not taken into account;
{\bf threshold nature of resources},
we assume that either an entity is available 
for all reactions,
or it is not available at all.

\begin{definition}[Reaction Result]
Given a (finite) set of entities $S$, and a  subset $W\subseteq S$, we define the following:
\begin{enumerate}
\item Let $a=(R,I,P)\in \mathit{rac}(S)$ be a reaction in $S$. 
The result of $a$ on $W$, denoted by $\mathit{res}_a(W)$, 
is defined by:
$$
\mathit{res}_a(W) \defeq
\left\{\begin{array}{ll}
P & \mbox{ if $\mathit{en}_a(W)$}\\
\emptyset & \mbox{ otherwise}
\end{array}
\right.
$$
where the enabling predicate is defined by
$\mathit{en}_a(W)\defeq R \subseteq W \wedge I \cap W = \emptyset$. 
\item Let $A\subseteq \mathit{rac}(S)$ be a finite set of reactions. The result of $A$ on $W$, denoted 
by $\mathit{res}_A(W)$,  is defined
by: $\mathit{res}_A(W) \defeq \bigcup_{a \in A} \mathit{res}_a(W)$.
\end{enumerate}
\end{definition}


Living cells are seen as open systems that  react with 
the external environment. 
The  behaviour of a RS is formalized in terms of 
{\em interactive processes}.

\begin{definition}[Interactive Process]
Let ${\cal A} = (S, A)$ be a RS and let $n \geq 0$. 
An $n$-steps
\emph{interactive process} in ${\cal A}$ is a pair $\pi = (\gamma, \delta)$ 
s.t.
%
$ \gamma=\{C_i\}_{i\in[0,n]}$ is the {\em context sequence}
and $\delta=\{D_i\}_{i\in[0,n]} $ is the {\em result sequence},
where 
$ C_{i}, D_{i}  \subseteq S$ for any $i\in[0,n]$, $D_0 = \emptyset$, and 
$D_{i+1} =  \mathit{res}_{A}(D_{i} \cup C_{i})$ for any $i \in [0,n-1]$.
We call $\tau = W_{0}, \ldots, W_{n}$ with $W_{i} \defeq C_{i} \cup D_{i}$, 
for any $i \in [0, n]$
the \emph{state sequence}.
\end{definition}


The context sequence $\gamma$ 
represents the environment.
The result sequence $\delta$ is entirely determined by $\gamma$ and $A$.
Each state $W_{i}$ in $\tau$
is the union of two sets: the context $C_{i}$
at step $i$ and the result set $D_i=\mathit{res}_{A}(W_{i-1})$ from the previous step.

%
%


%

Given a context sequence $\gamma$, we denote by $\gamma^k$ the shift of $\gamma$ starting at the $k$-th step. The shift notation will come in handy to draw a tight correspondence between the classic semantics of RS and the newly proposed SOS specification.

\begin{definition}[Sequence shift] 
Let
$ \gamma=\{C_i\}_{i\in[0,n]}$ a context sequence.
Given a positive integer $k\leq n$ we let 
$\gamma^k=\{C_{i+k}\}_{i\in[0,n-k]}$.
\end{definition}

We conclude this section with a simple example of RS.

\begin{example}\label{ex:ex1}
Here we consider a toy RS  defined as ${\cal A} = (S, A)$ where the set $S= \{\mathsf{a},\mathsf{b},\mathsf{c}\}$ only contains 
three entities, and the set of reactions $A=\{a_1\}$ only contains the reaction $a_1=(\{\mathsf{a},\mathsf{b}\},\{\mathsf{c}\},\{\mathsf{b}\})$, to be written more concisely as $(\mathsf{a}\mathsf{b}, \mathsf{c},\mathsf{b})$.
Then, we consider a $4\mathit{-}steps$ interactive process $\pi = (\gamma,\delta)$, where $\gamma = \{ C_0,C_1,C_2,C_3\} $, with $C_0 = \{\mathsf{a},\mathsf{b}\}$, $C_1 = \{ \mathsf{a}\}$, $C_2 = \{\mathsf{c}\}$, and $C_3 = \{\mathsf{c}\}$; and $\delta = \{D_0,D_1,D_2,D_3\}$, with $D_0 =\emptyset$, $D_1 = \{\mathsf{b}\}$, $D_2 = \{\mathsf{b}\}$, and $D_3 =\emptyset$.  
Then, the resulting state sequence is $$\tau = W_0, W_1, W_2, W_3= \{\mathsf{a},\mathsf{b}\}, \{\mathsf{a},\mathsf{b}\}, \{\mathsf{b},\mathsf{c}\}, \{\mathsf{c}\}.$$
In fact, it is easy to check that, e.g., $W_0 =C_0$, $D_1 = \mathit{res}_{A}(W_0) = \mathit{res}_{A}(\{\mathsf{a},\mathsf{b}\}) = \{\mathsf{b}\}$ because $\mathit{en}_a(W_0)$, and $W_1 = C_1\cup D_1 = \{ \mathsf{a}\}\cup \{ \mathsf{b}\} = \{\mathsf{a}, \mathsf{b}\}$.
\end{example}

\section{SOS Rules for Reaction Systems}
\label{sec:LTSforRS}

Inspired by classic process algebras, such as CCS~\cite{Milner80}, we introduce a syntax for RSs that resembles their original presentation and then equip each operator with some SOS inference rules that define its behaviour. This way: (1)~we establish a strong correspondence between terms of the signature and RSs; (2)~we derive an LTS semantics for each RS, where the states are terms, each transition corresponds to a step of the RS and transition labels retain some information needed for compositionality; (3)~we pave the way to the  RS enhancements  in Section~\ref{sec:extension}.

\begin{definition}[RS processes]\label{def:LTSforRS}
Let $S$ be a set of entities. An \emph{RS process} $\mathsf{P}$ is any term defined by the following grammar:
\[
\begin{array}{lcl}
\mathsf{P}&::=&[\mathsf{M}]\\
\mathsf{M} &::=&(R,I,P)~\mid~D~\mid \mathsf{K}~\mid~\mathsf{M}|\mathsf{M}\\
\mathsf{K}&::=& \nil~\mid~X~\mid~C.\mathsf{K}~\mid~\mathsf{K}+\mathsf{K}~\mid~\mathsf{rec}~X.~\mathsf{K}
\end{array}
\]
where $R,I,P\subseteq S$ are non empty sets of entities, $C,D\subseteq S$ are possibly empty set of entitities,  and $X$ is a process variable.
\end{definition}

An RS process  $\mathsf{P}$ embeds a \emph{mixture} process $\mathsf{M}$ obtained as the parallel composition of some reactions $(R,I,P)$, some set of currently present entities $D$ (possibly the empty set $\emptyset$), and some \emph{context} process $\mathsf{K}$.
We write $\prod_{i\in I} \mathsf{M}_i$ for the parallel composition of all $\mathsf{M}_i$ with $i\in I$. For example, $\prod_{i\in \{1,2\}} \mathsf{M}_i = \mathsf{M}_1~|~\mathsf{M}_2$.


A process context $\mathsf{K}$ is a possibly nondeterministic and recursive system: 
the nil context $\nil$ stops the computation;
the prefixed context $C.\mathsf{K}$ says that the entities in $C$ are immediately available to be consumed by the reactions, and then $\mathsf{K}$ is the context offered at the next step;
the non deterministic choice $\mathsf{K}_1+\mathsf{K}_2$ allows the context to behave either as $\mathsf{K}_1$ or $\mathsf{K}_2$;  
$X$ is a process variable, and $\mathsf{rec}~X.~\mathsf{K}$ is the usual recursive operator of process algebras.
We write $\sum_{i\in I} \mathsf{K}_i$ for the nondeterministic choice between all $\mathsf{K}_i$ with $i\in I$.

We say that $\mathsf{P}$ and $\mathsf{P}'$ are structurally equivalent, written $\mathsf{P} \equiv \mathsf{P}'$, when they denote the same term up to the laws of commutative monoids (unit, associativity and commutativity) for  parallel composition $\cdot | \cdot$, with $\emptyset$ as the unit, and the laws of idempotent and commutative monoids for  choice $\cdot +\cdot$, with $\nil$ as the unit. We also assume $D_1 | D_2 \equiv D_1\cup D_2$ for any $D_1,D_2\subseteq S$.

\begin{remark}
Note that the processes $\emptyset$ and $\nil$ are not interchangeable: as it will become clear from the operational semantics, the process $\emptyset$ can perform just a trivial transition to itself, while the process $\nil$ cannot perform any transition.
\end{remark}
 
\begin{definition}[RSs as RS processes]\label{def:fromRS}
Let ${\cal A} = (S, A)$  be  a RS, and $\pi = (\gamma, \delta)$ an $n$-step interactive process in ${\cal A}$, with $\gamma = \{C_i\}_{i\in[0,n]}$ and $\delta = \{D_i\}_{i\in[0,n]}$.
For any step $i\in[0,n]$, the corresponding RS process $\llbracket {\cal A},\pi \rrbracket_i$ is defined as follows:
$$
\llbracket {\cal A},\pi \rrbracket_i
\defeq 
\left[\prod_{a\in A} a~|~D_i~|~\mathsf{K}_{\gamma^i}\right]
$$
where the context process $\mathsf{K}_{\gamma^i} \defeq C_i.C_{i+1}.\cdots.C_n.\nil$ is the sequentialization of the entities offered by $\gamma^i$.
We write $\llbracket {\cal A},\pi \rrbracket$ as a shorthand for $\llbracket {\cal A},\pi \rrbracket_0$.
\end{definition}

\begin{example}\label{ex:ex2}
Here, we give the encoding of the reaction system, ${\cal A} = (S, A)$, defined in Example~\ref{ex:ex1}. The resulting RS process is as follows:
$$
\mathsf{P} = 
\llbracket {\cal A},\pi \rrbracket =   
\llbracket (\{\mathsf{a},\mathsf{b},\mathsf{c}\},\{(\mathsf{a}\mathsf{b}, \mathsf{c},\mathsf{b})\}),\pi \rrbracket 
=
[(\mathsf{a}\mathsf{b}, \mathsf{c},\mathsf{b})~|~ \emptyset~|~\mathsf{K}_{\gamma}]
\equiv
[(\mathsf{a}\mathsf{b}, \mathsf{c},\mathsf{b})~|~ \mathsf{K}_{\gamma}]
$$
where $\mathsf{K}_{\gamma} = \{\mathsf{a},\mathsf{b}\}.\{ \mathsf{a}\}. \{\mathsf{c}\}.\{\mathsf{c}\}.\nil$, written more concisely as $\mathsf{a}\mathsf{b}.\mathsf{a}.\mathsf{c}.\mathsf{c}.\nil$.
Note that $D_0=\emptyset$ is inessential and can be discarded thanks to structural congruence.
\end{example}

In Definition~\ref{def:fromRS} we have not exploited the entire potentialities of the syntax. In particular, the context $\mathsf{K}_\gamma$ is just a finite sequence of action prefixes induced by the set of entities provided by $\gamma$ at the various steps.
Our syntax allows for more general kinds of contexts as shown in the example below.
Nondeterminstic contexts can be used to collect several experiments, while recursion can be exploited to extract some regularity in the longterm behaviour of a Reaction System.
Together they offer any combination of in-breadth/in-depth analysis.

\begin{example}\label{ex:ex3}
Let us consider our running example. Suppose we want to enhance the behaviour of the context by defining a process $\mathsf{K}'= \mathsf{K}_1+\mathsf{K}_2$ that non-deterministically can behave as $ \mathsf{K}_1$ or as $\mathsf{K}_2$, where $ \mathsf{K}_1 = \mathsf{a}\mathsf{b}.\mathsf{a}.\mathsf{c}.\mathsf{c}.\nil$ (as in  Example~\ref{ex:ex2}), and $\mathsf{K}_2 = \mathsf{rec}~X.~\mathsf{a}\mathsf{b}.\mathsf{a}. X $ (which is a recursive behaviour that allows the reaction to be always enabled).
Then we simply define 
$
\mathsf{P}' \equiv [(\mathsf{a}\mathsf{b}, \mathsf{c},\mathsf{b})~|~\mathsf{K}']
$.
\end{example}

\begin{figure}[t]
$$
\infer[\scriptsize(\textit{Ent})]
{D \xrightarrow{\obs{D}{\emptyset,\emptyset,\emptyset}}\emptyset}
{}
\qquad
\infer[\scriptsize(\textit{Cxt})]
{C.\mathsf{K} \xrightarrow{\obs{C}{\emptyset,\emptyset,\emptyset}}\mathsf{K}}
{}
\qquad
\infer[\scriptsize(\textit{Rec})]
{\mathsf{rec}~X.~\mathsf{K} \xrightarrow{\obs{W}{R,I,P}}\mathsf{K}'}
{\mathsf{K}[^{\mathsf{rec}~X.~\mathsf{K}}/_{X}] \xrightarrow{\obs{W}{R,I,P}}\mathsf{K}'}
$$
\smallskip
$$
\infer[\scriptsize(\textit{Suml})]
{\mathsf{K}_1 + \mathsf{K}_2 \xrightarrow{\obs{W}{R,I,P}}\mathsf{K}'_1}
{\mathsf{K}_1 \xrightarrow{\obs{W}{R,I,P}}\mathsf{K}'_1}
\qquad
\infer[\scriptsize(\textit{Sumr})]
{\mathsf{K}_1 + \mathsf{K}_2 \xrightarrow{\obs{W}{R,I,P}}\mathsf{K}'_2}
{\mathsf{K}_2 \xrightarrow{\obs{W}{R,I,P}}\mathsf{K}'_2}
$$
\smallskip
$$
\infer[\scriptsize(\textit{Pro})]
{(R,I,P)  \xrightarrow{\obs{\emptyset}{R,I,P}}(R,I,P)~|~P}
{}
\qquad
\infer[\scriptsize(\textit{Inh})]
{(R,I,P)  \xrightarrow{\obs{\emptyset}{J,Q,\emptyset}}(R,I,P)}
{J \subseteq I & Q \subseteq R & J\cup Q\neq \emptyset}
$$
\smallskip
$$
\infer[\scriptsize(\textit{Par})]
{\mathsf{M}_1~|~\mathsf{M}_2\xrightarrow{\obs{W_1\cup W_2}{R_1\cup R_2,I_1\cup I_2,P_1\cup P_2}} \mathsf{M}'_1~|~\mathsf{M}'_2}
{\mathsf{M}_1 \xrightarrow{\obs{W_1}{R_1,I_1,P_1}} \mathsf{M}'_1 &
\mathsf{M}_2 \xrightarrow{\obs{W_2}{R_2,I_2,P_2}} \mathsf{M}'_2 &
(W_1\cup W_2 \cup R_1 \cup R_2) \cap (I_1 \cup I_2) = \emptyset}
$$
\smallskip
$$
\infer[\scriptsize(\textit{Sys})]
{[\mathsf{M}]\xrightarrow{\obs{W}{R,I,P}} [\mathsf{M}']}
{\mathsf{M}\xrightarrow{\obs{W}{R,I,P}} \mathsf{M}' &
R\subseteq W}
$$
\caption{SOS semantics of the reaction system processes.}
\label{fig:LTSforRS}
\end{figure}

\begin{definition}[Label]
A label is a tuple $\obs{W}{R,I,P}$  with $W,R,I,P\subseteq S$.
\end{definition}

In a transition label $\obs{W}{R,I,P}$, we record the set $W$ of entities currently in the system (produced in the previous step or provided by the context),  the set $R$ of  entities whose presence is assumed (either because they are needed as reactants on an applied reaction or because their presence prevents the application of some reaction); the set  $I$ of entities whose absence is assumed (either because they appear as inhibitors for an applied reaction or because their absence prevents the application of some reaction); the set $P$ of products of  all the applied reactions.

\begin{definition}[Operational semantics]
The operational semantics of processes is defined by the set of SOS inference rules in Figure~\ref{fig:LTSforRS}.
\end{definition}

The process $\nil$ has no transition. 
The rule $(\textit{Ent})$ makes available the entities in the (possibly empty) set $D$, then reduces to $\emptyset$.
As a special instance of $(\textit{Ent})$, $\emptyset \xrightarrow{\obs{\emptyset}{\emptyset,\emptyset,\emptyset}}\emptyset$.
The rule $(\textit{Cxt})$ says that a prefixed context process $C.\mathsf{K}$ makes available the entities  in the set $C$ and then reduces to $\mathsf{K}$. The rule $(\textit{Rec})$ is the classical rule for recursion.
Here, $\mathsf{K}[^{\mathsf{rec}~X.~\mathsf{K}}/_{X}]$ denotes the process obtained by replacing in $\mathsf{K}$ every free occurrence of the variable $X$ with its recursive definition $\mathsf{rec}~X.~\mathsf{K}$. For example $\mathsf{rec}~X.~\mathsf{a}.\mathsf{b}.X \xrightarrow{\obs{\mathsf{a}}{\emptyset,\emptyset,\emptyset}} \mathsf{b}.\mathsf{rec}~X.~\mathsf{a}.\mathsf{b}.X$
The rules $(\textit{Suml})$ and $(\textit{Sumr})$  select a move of either the left or the right component, resp., discarding the other process.
The rule $(\textit{Pro})$, executes the reaction $(R,I,P)$ (its reactants, inhibitors, and products
are recorded the label), which remains available at the next step together with $P$.
The rule $(\textit{Inh})$ applies when the reaction $(R,I,P)$ should not be executed;
it records in the label the possible causes for which the reaction is disabled: possibly some inhibiting entities $(J \subseteq I)$ are present or some reactants $(Q \subseteq R)$ are missing, with $J \cup Q \neq \emptyset$, as at least one cause is needed for explaining why the reaction is not enabled.\footnote{Conceptually, one could extend labels to record $J$ and $Q$ in separate positions from $R$ and $I$, respectively, like in $\obs{W}{R,J,I,Q,P}$. However, one would then need to rewrite the side conditions of all the rules by replacing $R$ with $R\cup J$ and $I$ with $I\cup Q$, because the distinction is never exploited in the SOS rules.}
The rule $(\textit{Par})$ puts two processes in parallel by pooling their labels and joining all the set components of the labels; a sanity check is required to guarantee that there is no conflict between reactants and inhibitors of the applied reactions.
Finally, the rule $(\textit{Sys})$ requires that all the processes of the systems have been considered, and 
also checks that all the needed reactants are actually available in the system ($R \subseteq W$). In fact this constraint can only be met on top of all processes. The check that inhibitors are absent ($I\cap W = \emptyset$) is not necessary, as it is embedded in rule $(\textit{Par})$.

\begin{example}\label{ex:ex4}
Let us consider the RS process $\mathsf{P}_0\defeq [(\mathsf{a}\mathsf{b}, \mathsf{c},\mathsf{b})~|~\mathsf{a}\mathsf{b}.\mathsf{a}.\mathsf{c}.\mathsf{c}.\nil]$ from Example~\ref{ex:ex2}.
The process $\mathsf{P}_0$ has a unique outgoing transition, whose formal derivation is given below:
$$
\infer[\scriptsize(\textit{Sys})]
{[(\mathsf{a}\mathsf{b}, \mathsf{c},\mathsf{b})~|~\mathsf{a}\mathsf{b}.\mathsf{a}.\mathsf{c}.\mathsf{c}.\nil] \xrightarrow{\obs{\mathsf{a}\mathsf{b}}{\mathsf{a}\mathsf{b}, \mathsf{c},\mathsf{b}}} [(\mathsf{a}\mathsf{b}, \mathsf{c},\mathsf{b})~|~\mathsf{b}~|~\mathsf{a}.\mathsf{c}.\mathsf{c}.\nil]}
{\infer[\scriptsize(\textit{Par})]
          {(\mathsf{a}\mathsf{b}, \mathsf{c},\mathsf{b})~|~\mathsf{a}\mathsf{b}.\mathsf{a}.\mathsf{c}.\mathsf{c}.\nil \xrightarrow{\obs{\mathsf{a}\mathsf{b}}{\mathsf{a}\mathsf{b}, \mathsf{c},\mathsf{b}}} (\mathsf{a}\mathsf{b}, \mathsf{c},\mathsf{b})~|~\mathsf{b}~|~\mathsf{a}.\mathsf{c}.\mathsf{c}.\nil}
          {\infer[\scriptsize(\textit{Pro})]
                   {(\mathsf{a}\mathsf{b}, \mathsf{c},\mathsf{b}) \xrightarrow{\obs{\emptyset} {\mathsf{a}\mathsf{b}, \mathsf{c},\mathsf{b}}} (\mathsf{a}\mathsf{b}, \mathsf{c},\mathsf{b})~|~\mathsf{b}}
                   {}
         &\infer[\scriptsize(\textit{Cxt})]
                   {\mathsf{a}\mathsf{b}.\mathsf{a}.\mathsf{c}.\mathsf{c}.\nil \xrightarrow{\obs{\mathsf{a}\mathsf{b}}{\emptyset,\emptyset,\emptyset}} \mathsf{a}.\mathsf{c}.\mathsf{c}.\nil}
                   {}}
}
$$
The target process $\mathsf{P}_1\defeq [(\mathsf{a}\mathsf{b}, \mathsf{c},\mathsf{b})~|~\mathsf{b}~|~\mathsf{a}.\mathsf{c}.\mathsf{c}.\nil]$ has also a unique outgoing transition, namely:
$$
\mathsf{P}_1 = [(\mathsf{a}\mathsf{b}, \mathsf{c},\mathsf{b})~|~\mathsf{b}~|~\mathsf{a}.\mathsf{c}.\mathsf{c}.\nil] \xrightarrow{\obs{\mathsf{a}\mathsf{b}}{\mathsf{a}\mathsf{b}, \mathsf{c},\mathsf{b}}} [(\mathsf{a}\mathsf{b}, \mathsf{c},\mathsf{b})~|~\mathsf{b}~|~\mathsf{c}.\mathsf{c}.\nil] =\mathsf{P}_2
$$
Instead the process $\mathsf{P}_2$
has three outgoing transitions, each providing a different justification to the fact that the reaction  $(\mathsf{a}\mathsf{b}, \mathsf{c},\mathsf{b})$ is not enabled:
\begin{enumerate}
\item $[(\mathsf{a}\mathsf{b}, \mathsf{c},\mathsf{b})~|~\mathsf{b}~|~\mathsf{c}.\mathsf{c}.\nil] \xrightarrow{\obs{\mathsf{b}\mathsf{c}}{\mathsf{c}, \mathsf{a},\emptyset}} [(\mathsf{a}\mathsf{b}, \mathsf{c},\mathsf{b})~|~\mathsf{c}.\nil]$, where the label shows that the presence of $\mathsf{c}$ and the absence of $\mathsf{a}$ inhibit the reaction;
\item $[(\mathsf{a}\mathsf{b}, \mathsf{c},\mathsf{b})~|~\mathsf{b}~|~\mathsf{c}.\mathsf{c}.\nil] \xrightarrow{\obs{\mathsf{b}\mathsf{c}}{\mathsf{c}, \emptyset,\emptyset}} [(\mathsf{a}\mathsf{b}, \mathsf{c},\mathsf{b})~|~\mathsf{c}.\nil]$, where it is only observed that the presence of $\mathsf{c}$  has played some role in inhibiting the reaction;
\item $[(\mathsf{a}\mathsf{b}, \mathsf{c},\mathsf{b})~|~\mathsf{b}~|~\mathsf{c}.\mathsf{c}.\nil] \xrightarrow{\obs{\mathsf{b}\mathsf{c}}{ \emptyset,\mathsf{a},\emptyset}} [(\mathsf{a}\mathsf{b}, \mathsf{c},\mathsf{b})~|~\mathsf{c}.\nil]$, where it is only observed that the absence of $\mathsf{a}$  has played some role in inhibiting the reaction.
\end{enumerate}
Notably, the three transitions have the same target process 
$\mathsf{P}_3\defeq [(\mathsf{a}\mathsf{b}, \mathsf{c},\mathsf{b})~|~\mathsf{c}.\nil]$.

Finally, the process $\mathsf{P}_3$ has seven transitions all leading to $\mathsf{P}_4\defeq [(\mathsf{a}\mathsf{b}, \mathsf{c},\mathsf{b})~|~\nil]$. Their labels are of the form $\obs{c}{J,Q,\emptyset}$ with $J\subseteq \mathsf{c}$, $Q\subseteq\mathsf{a}\mathsf{b}$ and $J\cup Q\neq \emptyset$. Each label provides a different explanation why the reaction is not enabled.
\end{example}


The following technical lemmas express some relevant properties of the transition system and can be proved by straightforward rule induction.

\begin{lemma}\label{lemma1}
If $\mathsf{M} \xrightarrow{\obs{W}{R,I,P}} \mathsf{M}'$ then $\mathsf{M}' \equiv \mathsf{M}''| P$ for some $\mathsf{M}''$.
\end{lemma}

\begin{lemma}\label{lemma2}
If $\prod_{a\in A} a \xrightarrow{\obs{W}{R,I,P}} \mathsf{M}$ then $W=\emptyset$ and $\mathsf{M} \equiv \prod_{a\in A} a~|~P$.
\end{lemma}

\begin{lemma}\label{lemma3}
If $\mathsf{M} \xrightarrow{\obs{W}{R,I,P}} \mathsf{M}'$ then $(W\cup R)\cap I = \emptyset$.
\end{lemma}

\begin{lemma}\label{lemma4}
If $\mathsf{P} \xrightarrow{\obs{W}{R,I,P}} \mathsf{P}'$ then $R\subseteq W$ and $W\cap I=\emptyset$.
\end{lemma}

The main theorem shows that the rewrite steps of a RS exactly match  the transitions of its corresponding RS process.

\begin{restatable}{theorem}{theoCorrespondence}
\label{the:corr}
Let ${\cal A} = (S, A)$  be  a RS, and $\pi=(\gamma,\delta)$ an $n$-step interactive process in ${\cal A}$ with $\gamma = \{C_i\}_{i\in[0,n]}$, $\delta = \{D_i\}_{i\in[0,n]}$, and let
$W_i \defeq C_i\cup D_i$ and $\mathsf{P}_i \defeq \llbracket {\cal A},\pi \rrbracket_i$ for any $i\in[0,n]$.
Then:
\begin{enumerate}
\item$\forall i\in[0,n-1]$,
$\mathsf{P}_i \xrightarrow{\obs{W}{R,I,P}} \mathsf{P}$
 implies
 $W= W_i$, $P= D_{i+1}$ and
$\mathsf{P} \equiv \mathsf{P}_{i+1}$;
\item $\forall i\in[0,n-1]$, there exists
$R,I \subseteq S$ such that 
$\mathsf{P}_i \xrightarrow{\obs{W_i}{R,I,D_{i+1}}} \mathsf{P}_{i+1}$.
\end{enumerate}
\end{restatable}

\begin{remark}
Note that the process $\mathsf{P}_n = \llbracket {\cal A},\pi \rrbracket_n = [\prod_{a\in A}a~|~D_n~|~C_n.\nil]$ has one more transition available (the $(n+1)$-th step from $\mathsf{P}_0$), even if the standard theory of RSs stops the computation after $n$ steps. We thus have additional steps
$$
\mathsf{P}_n \xrightarrow{\obs{W_n}{R_n,I_n,\mathit{res}_A(W_n)}} 
\left[\prod_{a\in A}a~|~\mathit{res}_A(W_n)~|~\nil\right]
$$
for suitable $R_n,I_n\subseteq S$.
The target process contains $\nil$ and therefore is deadlock.
\end{remark}

Example~\ref{ex:ex4} shows that we can have redundant transitions because of rule $(\textit{Inh})$. However, they can be easily detected and eliminated by considering a notion of dominance. To this aim we introduce an order relation $\sqsubseteq$ over pairs of set of entities defined as follows:
$$
(R',I') \sqsubseteq (R,I)\quad\mbox{if}\quad R'\subseteq R\wedge I'\subseteq I.
$$

\begin{definition}[Dominance]
A transition $\mathsf{P} \xrightarrow{\obs{W}{R',I',P}} \mathsf{P}'$ is \emph{dominated} if there exists another transition $\mathsf{P} \xrightarrow{\obs{W}{R,I,P}} \mathsf{P}'$ such that $(R',I') \sqsubset (R,I)$. 
\end{definition}

Note that in the definition of dominance we require the dominated transition to have the same source and target processes as the dominant transition, and that their labels carry also the same sets $W$ and $P$.

Finally, we can immediately derive an LTS, whose transitions are written using double arrows, where only dominant transitions are considered. The LTS is defined by the additional SOS rule $(\textit{Dom})$ below:
$$
\infer[\scriptsize(\textit{Dom})]
{\mathsf{P}\xRightarrow{\obs{W}{R,I,P}} \mathsf{P}'}
{\mathsf{P}\xrightarrow{\obs{W}{R,I,P}} \mathsf{P}' &
(R,I) = \max_{\sqsubseteq} \{(R',I')~|~\mathsf{P}\xrightarrow{\obs{W}{R',I',P}} \mathsf{P}'\}}
$$

In other words, a transition $\mathsf{P}\xRightarrow{\obs{W}{R,I,P}} \mathsf{P}'$ guarantees that any instance of the rule $(\textit{Inh})$ is applied in a way that maximizes the sets $J$ and $Q$ (given the overall available entities $W$).

\begin{example}
Looking back at Example~\ref{ex:ex4}, both transitions 
$\mathsf{P}_2
\xrightarrow{\obs{\mathsf{b}\mathsf{c}}{\mathsf{c}, \emptyset,\emptyset}} 
\mathsf{P}_3$
and
$\mathsf{P}_2 
\xrightarrow{\obs{\mathsf{b}\mathsf{c}}{ \emptyset,\mathsf{a},\emptyset}} 
\mathsf{P}_3$
are dominated by 
$\mathsf{P}_2
\xrightarrow{\obs{\mathsf{b}\mathsf{c}}{\mathsf{c}, \mathsf{a},\emptyset}} 
\mathsf{P}_3$.
Therefore, the process $\mathsf{P}_2 = [(\mathsf{a}\mathsf{b}, \mathsf{c},\mathsf{b})~|~\mathsf{b}~|~\mathsf{c}.\mathsf{c}.\nil]$ has a unique (double-arrow) transition
$\mathsf{P}_2 
\xRightarrow{\obs{\mathsf{b}\mathsf{c}}{\mathsf{c}, \mathsf{a},\emptyset}}
\mathsf{P}_3$.
\end{example}

\section{Bio-simulation}
\label{sec:biosimulation}

Bisimulation equivalences~\cite{Sangiorgi:introbis} play a central role in process algebras.
They can be defined in terms of coinductive games, of fixpoint theory and of logics.
The bisimulation game is played by an attacker and a defender: the former wants to disprove the equivalence between two processes $p$ and $q$, the latter that $p$ and $q$ are equivalent. The game is turn based: at each turn the attacker picks one process, e.g., $p$,  and one transition $p\xrightarrow{\lambda} p'$ and the defender must reply by picking one transition $q\xrightarrow{\lambda} q'$ of the other process with exactly the same label $\lambda$; then the game continues challenging the equivalence between $p'$ and $q'$.
The game ends when the attacker has no transition available, and the defender wins, or when defender cannot match the move of the attacker, and the attacker wins. The defender also wins  if the game doesn't end. Then $p$ and $q$ are not equivalent iff the attacker has a winning strategy.
There are many variants of the bisimulation for process algebras,  for example the
barbed bisimulation~\cite{10.1007/3-540-55719-9_114} only considers the execution of invisible actions, and then equates two processes when they expose the same prefixes; for the mobile ambients~\cite{CardelliG00}, a process algebra equipped with a reduction semantics, a notion of
behavioural equivalence equates two processes when they expose the same ambients~\cite{GC03}. 

 In the case of biological systems, the classical notion of bisimulation can be too concrete.
In fact, in a biological soup, a high number of interactions occur every time instant, and generally, biologists
are only interested to analyse a small subset of them and to focus on a subset of entities.
In the case of RS processes, the labels that we used for the LTS consider too many details and convey too much information: they record the entire information about all the reactions that have been applied in one transition,  the entities that acted as reactants, as inhibitors or as products, or that were available in the state.
All this information  stored in the label is necessary to compose a transition in a modular way. 
Depending on the application, only a suitable abstraction over the label can be of interest.
For this reason, following the approach introduced in Brodo et al.~\cite{TCS:RSLink}, we propose an alternative notion of bisimulation, called  \emph{bio-simulation}, that compares two biological systems by restricting the observation to only a limited set of events that are of particular interest.
With respect to the work in Brodo et al.~\cite{TCS:RSLink}, here the labels are easier to manage and simpler to parse.

In a way, at each step of the bisimulation game, we want to query our labels about some partial information.
To this goal, we define an  assertion language
 to express detailed and partial queries about what happened in a single transition.

\begin{example} \label{properties_ex}
For instance we would like to express properties about each step of the bio-simulation of a system like the ones below:
\begin{enumerate}
\item Has the presence of the entity $\mathsf{a}$ been exploited by some reaction?
\item Have the entities $\mathsf{a}$ and $\mathsf{b}$ been produced by some reaction?
\item Have the entities $\mathsf{a}$ or $\mathsf{c}$ been provided by the state? 
\item Has the reaction $(\mathsf{a}\mathsf{b},\mathsf{c},\mathsf{b})$  been applied or not? 
\end{enumerate}
\end{example}

As detailed before, in the following we assume that the context can be non-deterministic, otherwise it makes little sense to rely on bisimulation to observe the branching structure of system dynamics.

The bio-simulation approach works as follows: first we introduce an assertion language to abstract away some information from the labels; then we define a bisimilarity equivalence that is parametric to a given assertion, called bio-similarity; finally we give a logical characterisation of bio-similarity, called bio-logical equivalence, by tailoring the classical HML to the given assertion.

\subsection{Assertion language}

An assertion is a formula that predicates on the labels of our LTS.
The assertion language that we propose is very basic, but can be extended if necessary.

%

\begin{definition}[Assertion Language]
\label{def:assetionl}
Given a set of entities $S$, assertions $\mathsf{F}$ on $S$ are built from the following syntax, where  $E\subseteq S$ and $\mathit{Pos} \in\{ \mathcal{W}, \mathcal{R}, \mathcal{I}, \mathcal{P}\}$:
$$
\begin{array}{lcl}
\mathsf{F} & ::= & 
E  \subseteq \mathit{Pos}
~\mid~ ?  \in \mathit{Pos}  ~\mid ~ 
\mathsf{F} \vee \mathsf{F} ~\mid ~ 
\mathsf{F} \wedge \mathsf{F}~\mid ~
\mathsf{F}  ~\widehat{\ }~\mathsf{F} ~\mid
\neg \mathsf{F}

\end{array}
$$
\end{definition}


Roughly,  $\mathit{Pos}$  distinguishes different positions in the labels:
$\mathcal{W}$ stands for entities provided by current state,
$\mathcal{R}$ stands for reactants,
$\mathcal{I}$ stands for inhibitors, and
$\mathcal{P}$ stands for products.
An assertion $\mathsf{F}$ is either the membership of a subset of entities $E$ in a given position $\mathit{Pos}$, $E\subseteq \mathit{Pos}$,
the test of $\mathit{Pos}$ for non-emptyness, $?\in \mathit{Pos}$, 
the disjunction of two assertions $\mathsf{F}_1 \vee \mathsf{F}_2$, their conjunction $\mathsf{F}_1 \wedge \mathsf{F}_2$,  their exclusive or $\mathsf{F}_1~\widehat{\ }~\mathsf{F}_2$,
or the negation of an assertion $\neg  \mathsf{F}$.


\begin{definition}[Satisfaction of Assertion]\label{def:sati}
Let  $\upsilon= \obs{W}{R,I,P}$ be  a transition label, and $\mathsf{F}$ be an assertion.
We write $\upsilon \entails \mathsf{F}$ (read as the transition label $\upsilon$ satisfies the assertion $\mathsf{F}$)  if and only if the following hold:
\[
\begin{array}{lcl}
\upsilon \entails E \subseteq \mathit{Pos} & \mbox{ iff  } & E\subseteq \mathsf{select}(\upsilon,\mathit{Pos}) \\ 
\upsilon \entails ? \in \mathit{Pos} &  \mbox{ iff  }  &  \mathsf{select}(\upsilon,\mathit{Pos}) \neq \emptyset \\ 
\upsilon \entails  \mathsf{F}_1 \wedge \mathsf{F}_2 &  \mbox{ iff  }  &\upsilon \entails  \mathsf{F}_1 \wedge  \upsilon \entails  \mathsf{F}_2 \\
\upsilon \entails  \mathsf{F}_1 \vee \mathsf{F}_2 &  \mbox{ iff  } &\upsilon \entails  \mathsf{F}_1 \vee  \upsilon \entails  \mathsf{F}_2 \\
\upsilon \entails  \mathsf{F}_1~\widehat{\ }~\mathsf{F}_2 &  \mbox{ iff  } &( \upsilon \entails   \mathsf{F}_1 \wedge  \upsilon \entails \neg \mathsf{F}_2 ) \vee (\upsilon \entails   \neg \mathsf{F}_1 \wedge  \upsilon \entails  \mathsf{F}_2 ) \\
\upsilon \entails  \neg \mathsf{F} &  \mbox{ iff  }  &\upsilon \not\entails  \mathsf{F}
\end{array}
\]
$$
 \text{where}\qquad\mathsf{select}(\obs{W}{R,I,P},\mathit{Pos}) \defeq \left\{
 \begin{array}{ll}
 W & \mbox{ if $\mathit{Pos} = \mathcal{W}$}\\
 R & \mbox{ if $\mathit{Pos} = \mathcal{R}$}\\
 I & \mbox{ if $\mathit{Pos} = \mathcal{I}$}\\
 P & \mbox{ if $\mathit{Pos} = \mathcal{P}$}
 \end{array}
 \right.
 $$
\end{definition}

\noindent
Given two transition labels $v,w$ we write $v\equiv_{\mathsf{F}} w$ if $v \entails \mathsf{F}\ \Leftrightarrow\ w \entails \mathsf{F}$, i.e. if both $v,w$ satisfy $\mathsf{F}$ or they both do not.

\begin{example}\label{ex:assertions}
Some assertions matching the queries listed in Example~\ref{properties_ex} are:
\begin{enumerate}
\item $\mathsf{F}_1 \defeq \mathsf{a}\subseteq \mathcal{R}$
\item $\mathsf{F}_2 \defeq \mathsf{a}\mathsf{b}\subseteq \mathcal{P}$
\item $\mathsf{F}_3 \defeq \mathsf{a}\subseteq \mathcal{W} \vee \mathsf{c}\subseteq \mathcal{W}$
\item $\mathsf{F}_4 \defeq \mathsf{ab}\subseteq \mathcal{R} \wedge \mathsf{c}\subseteq \mathcal{I}$ checks if the reaction has been applied, while $\mathsf{F}_5 \defeq \mathsf{a}\subseteq \mathcal{I} \vee \mathsf{b}\subseteq \mathcal{I} \vee \mathsf{c}\subseteq \mathcal{R}$ the opposite case. Alternatively, we can set $\mathsf{F}_5 \defeq \neg \mathsf{F}_4$.
\end{enumerate}
If we take the label $\upsilon=\obs{\mathsf{a}\mathsf{b}}{\mathsf{a}\mathsf{b}, \mathsf{c},\mathsf{b}}$ it is immediate to check that
$$
\upsilon \models \mathsf{F}_1 
\qquad
\upsilon \not\models \mathsf{F}_2
\qquad
\upsilon \models \mathsf{F}_3 
\qquad
\upsilon \models \mathsf{F}_4 
\qquad
\upsilon \not\models \mathsf{F}_5 
$$
\end{example}

 
 With respect to the assertion language proposed in our previous paper~\cite{TCS:RSLink}, the new one has less expressive power
as it is not possible  to immediately distinguish the reagents, the inhibitors and the products referred to each reaction applied, or to know the reason why a reaction has not been applied.
However, these informations can be retrieved by the reaction definition.
The main interest  of this proposal is that it is directly applied to the LTS tailored for RSs.

\subsection{Bio-similarity and bio-logical equivalence}

The notion of bio-simulation builds on the above language of assertions to parameterize the induced equivalence on the property of interest. 
Please recall that we have defined the behaviour of the context in a non deterministic way, thus 
at each step, different possible sets of entities can be provided to the system and different sets of reaction can be enabled/disabled. 
Bio-simulation can thus be used to compare the behaviour of different systems that share some of the reactions or entities or also to compare the behaviour of the same set of reaction rules when different contexts are provided.

\begin{definition}[Bio-similarity  $ \sim_{\mathsf{F}}$~\cite{TCS:RSLink}]
Given an assertion $\mathsf{F}$, a \emph{bio-simulation} $\mathbf{R}_{\mathsf{F}}$ that respects $\mathsf{F}$ is a binary relation over RS  processes s.t., if $\mathsf{P}~\mathrel{\mathbf{R}_{\mathsf{F}}}~\mathsf{Q}$ then:
 \begin{itemize} 
 \item
 $\forall \upsilon,\mathsf{P}'$ s.t. 
 $\mathsf{P} \xRightarrow{\upsilon} \mathsf{P}'$, then $\exists w,\mathsf{Q}'$ s.t. $\mathsf{Q}  \xRightarrow{w} \mathsf{Q}'$ with $\upsilon\equiv_{\mathsf{F}} w$ and $\mathsf{P}' \mathrel{\mathbf{R}_{\mathsf{F}}} \mathsf{Q}'$.
\item
 $\forall w,\mathsf{Q}'$ s.t. 
 $\mathsf{Q} \xRightarrow{w} \mathsf{Q}'$, then $\exists \upsilon,\mathsf{P}'$ s.t. $\mathsf{P}  \xRightarrow{\upsilon} \mathsf{P}'$ with $\upsilon\equiv_{\mathsf{F}} w$ and $\mathsf{P}' \mathrel{\mathbf{R}_{\mathsf{F}}} \mathsf{Q}'$.
\end{itemize}

We let $ \sim_{\mathsf{F}} $ denote the largest  bio-simulation and we say that $\mathsf{P}$ is \emph{bio-similar} to $\mathsf{Q}$, with respect to $\mathsf{F}$, if $\mathsf{P} \sim_{\mathsf{F}} \mathsf{Q}$.
\end{definition}

\begin{remark}
An alternative way to look at a bio-simulation that respects $\mathsf{F}$ is to define it as an ordinary bisimulation over the transition system labelled over $\{\mathsf{F},\neg \mathsf{F}\}$ obtained by transforming each transition $\mathsf{P} \xRightarrow{\upsilon} \mathsf{P}'$ such that $\upsilon\entails \mathsf{F}$ into $\mathsf{P} \xRightarrow{\mathsf{F}} \mathsf{P}'$ and  each transition $\mathsf{P} \xRightarrow{\upsilon} \mathsf{P}'$ such that $\upsilon\not\entails \mathsf{F}$ into $\mathsf{P} \xRightarrow{\neg \mathsf{F}} \mathsf{P}'$.
\end{remark}

It can be easily shown that the identity relation is a bio-simulation and that bio-simulations are closed under (relational) inverse, composition and union and that, as a consequence, bio-similarity is an equivalence relation.

\begin{example}\label{ex:biobis}
Let us consider some variants of our working example.  
The behavior of $\mathsf{P}_0\defeq [(\mathsf{a}\mathsf{b}, \mathsf{c},\mathsf{b})~|~\mathsf{a}\mathsf{b}.\mathsf{a}.\mathsf{a}\mathsf{c}.\nil]$ is deterministic, and its  unique trace of labels is:
$$
\xymatrix@C=40pt{
{\mathsf{P}_0}\ar@{=>}[r]^{\obs{\mathsf{a}\mathsf{b}}{\mathsf{a}\mathsf{b}, \mathsf{c},\mathsf{b}}} &
{\mathsf{P}_1} \ar@{=>}[r]^{\obs{\mathsf{a}\mathsf{b}}{\mathsf{a}\mathsf{b}, \mathsf{c},\mathsf{b}}} &
{\mathsf{P}_2} \ar@{=>}[r]^(.3){\obs{\mathsf{a}\mathsf{b}\mathsf{c}}{\mathsf{c}, \emptyset,\emptyset}} &
{[(\mathsf{a}\mathsf{b}, \mathsf{c},\mathsf{b})|\nil]}
}
$$
 Instead, the behavior of $\mathsf{P}'_0\defeq [(\mathsf{a}\mathsf{b}, \mathsf{c},\mathsf{b})~|~(\mathsf{a}\mathsf{b}.\mathsf{a}.\mathsf{a}\mathsf{c}.\nil +\mathsf{a}\mathsf{b}.\mathsf{a}.\mathsf{a}.\nil )]$ is non deterministic. Now there are two possible  traces of labels: the first trace is equal to the above one, and the other one follows:
 $$
\xymatrix@R=5pt@C=40pt{
{\mathsf{P}'_0}
\ar@{=>}[r]^{\obs{\mathsf{a}\mathsf{b}}{\mathsf{a}\mathsf{b}, \mathsf{c},\mathsf{b}}} 
\ar@{=>}[rd]_{\obs{\mathsf{a}\mathsf{b}}{\mathsf{a}\mathsf{b}, \mathsf{c},\mathsf{b}}} &
{\mathsf{P}_1} \ar@{=>}[r]^{\obs{\mathsf{a}\mathsf{b}}{\mathsf{a}\mathsf{b}, \mathsf{c},\mathsf{b}}} &
{\mathsf{P}_2} \ar@{=>}[r]^(.3){\obs{\mathsf{a}\mathsf{b}\mathsf{c}}{\mathsf{c}, \emptyset,\emptyset}} &
{[(\mathsf{a}\mathsf{b}, \mathsf{c},\mathsf{b})|\nil]}\\
&{\mathsf{P}'_1} \ar@{=>}[r]^{\obs{\mathsf{a}\mathsf{b}}{\mathsf{a}\mathsf{b}, \mathsf{c},\mathsf{b}}}
&{\mathsf{P}'_2} \ar@{=>}[r]^(.3){\obs{\mathsf{a}\mathsf{b}}{\mathsf{a}\mathsf{b}, \mathsf{c},\mathsf{b}}}&
{[(\mathsf{a}\mathsf{b}, \mathsf{c},\mathsf{b})|\mathsf{b}|\nil]}
}
 $$ 
Now, it is easy to check that the two processes $\mathsf{P}_0$, $\mathsf{P}'_0$ are not bio-similar w.r.t.
the assertion $\mathsf{F}_1\defeq \mathsf{c} \in  \mathcal{E}$, requiring that in the state configuration entity $\mathsf{c}$ is present, and are bio-similar w.r.t. the assertion 
 $\mathsf{F}_2\defeq (\mathsf{a} \in  \mathcal{R})~\widehat{\ }~(\mathsf{c} \in  \mathcal{R})$, requiring that either $\mathsf{c}$ or $\mathsf{a}$ are used as reactants.
\end{example}

Now, we introduce a slightly modified version of the Hennessy-Milner Logic~\cite{HM80}, called bioHML; due to the reasons we explained above, we do not want to look at the complete transition labels, thus we rely on our simple assertion language to make it parametric to the assertion $\mathsf{F}$ of interest:

\begin{definition}[BioHML~\cite{TCS:RSLink}]
Let   $\mathsf{F}$ be an assertion, then 
the set of bioHML formulas $\mathsf{G}$ that respects $\mathsf{F}$ are built by the following syntax, where $\chi \in\{ \mathsf{F},\neg \mathsf{F}\}$:
$$
\begin{array}{rcl}
\mathsf{G},\mathsf{H} &::= &{\tt t} ~\mid~ {\tt f} ~\mid~ \mathsf{G}\wedge \mathsf{G} ~\mid~ \mathsf{G} \vee \mathsf{G} ~\mid~ \langle \chi \rangle \mathsf{G} ~\mid~ [\chi]\mathsf{G}
\end{array}
$$
\end{definition}

\begin{remark}
An alternative way to look at bioHML formulas is as ordinary HML formulas over the set of labels $\{\mathsf{F},\neg \mathsf{F}\}$.
\end{remark}

The semantics of a bioHML formula is the set of processes that satisfy it.

\begin{definition}[Semantics of BioHML]
Let $\mathbb{P}$ denote the set of all RS processes over $S$.
For a BioHML formula $\mathsf{G}$, we define $\llbracket \mathsf{G}\rrbracket \subseteq \mathbb{P}$ inductively on $\mathsf{G}$:
\[
\begin{array}{ccl@{\hspace{1cm}}ccl@{\hspace{1cm}}ccl@{\hspace{1cm}}ccl}
 \llbracket {\tt t}\rrbracket &\defeq& \mathbb{P} &  
 \llbracket {\tt f}\rrbracket& \defeq& \emptyset &
\llbracket \mathsf{G}\wedge \mathsf{H}\rrbracket &\defeq&  \llbracket \mathsf{G}\rrbracket  \cap \llbracket \mathsf{H}\rrbracket  &
  \llbracket \mathsf{G}\vee \mathsf{H}\rrbracket &\defeq&  \llbracket \mathsf{G}\rrbracket  \cup \llbracket \mathsf{H}\rrbracket 
\end{array}
\]
\[
\begin{array}{lcl}
 \llbracket  \langle \chi \rangle \mathsf{G} \rrbracket &\defeq& \{\mathsf{P} \in \mathbb{P}: \exists \upsilon,\mathsf{P}'.\ \mathsf{P}\xRightarrow{\upsilon}\mathsf{P}' \mbox{ with } \upsilon \entails \chi \mbox{ and } \mathsf{P}' \in  \llbracket  \mathsf{G} \rrbracket\} \\
 \llbracket [\chi]\mathsf{G} \rrbracket &\defeq& \{\mathsf{P} \in \mathbb{P}: \forall \upsilon,\mathsf{P}'.\ \mathsf{P}\xRightarrow{\upsilon}\mathsf{P}' \mbox{ implies } \upsilon \entails \chi \mbox{ and } \mathsf{P}' \in  \llbracket  \mathsf{G} \rrbracket\}
\end{array}
\]

We write $\mathsf{P} \entails \mathsf{G}$ ($\mathsf{P}$ satisfies $\mathsf{G}$) if and only if $\mathsf{P} \in  \llbracket \mathsf{G} \rrbracket$.
\end{definition}

Negation is not included in the syntax, but the converse $\overline{\mathsf{G}}$ of a bioHML formula $\mathsf{G}$ can be easily defined inductively in the same way as for HML logic.

%

We let $\mathcal{L}_{\mathsf{F}}$ be the set of all bioHML formulas that respects $\mathsf{F}$.

\begin{definition}[Bio-logical equivalence]
We say that $\mathsf{P},\mathsf{Q}$ are \emph{bio-logically} equivalent w.r.t. $\mathsf{F}$, written $\mathsf{P} \equiv_{\mathcal{L}_{\mathsf{F}}}\mathsf{Q}$, when $\mathsf{P}$ and $\mathsf{Q}$ satisfy the exactly the same bioHML formulas in $\mathcal{L}_{\mathsf{F}}$, i.e. when for any $\mathsf{G}\in \mathcal{L}_{\mathsf{F}}$ we have $\mathsf{P} \entails \mathsf{G}\ \Leftrightarrow \mathsf{Q} \entails \mathsf{G}$.
\end{definition}

Finally, we extend the classical result establishing the correspondence between the logical equivalence induced by HML with bisimilarity for proving that bio-similarity coincides with bio-logical equivalence.

\begin{restatable}[Correspondence~\cite{TCS:RSLink}]{theorem}{theocorr}
$\sim_{\mathsf{F}}\ =\ \equiv_{\mathcal{L}_{\mathsf{F}}}$
\end{restatable}

\begin{example}
We continue by considering our running example in Example~\ref{ex:biobis}.
 There already is the evidence that the two processes  $\mathsf{P}_0\defeq [(\mathsf{a}\mathsf{b}, \mathsf{c},\mathsf{b})~|~\mathsf{a}\mathsf{b}.\mathsf{a}.\mathsf{a}\mathsf{c}.\nil]$ and 
$\mathsf{P}'_0\defeq [(\mathsf{a}\mathsf{b}, \mathsf{c},\mathsf{b})~|~(\mathsf{a}\mathsf{b}.\mathsf{a}.\mathsf{a}\mathsf{c}.\nil +\mathsf{a}\mathsf{b}.\mathsf{a}.\mathsf{a}.\nil )]$ are not bio-similar w.r.t. the assertion 
 $\mathsf{F}_1\defeq \mathsf{c} \in  \mathcal{W}$. Here, we give a  bioHML formula that distinguishes $\mathsf{P}_0$
and $\mathsf{P}_0'$:
$$\mathsf{G} \defeq\langle \neg\mathsf{F}_1 \rangle  [ \neg\mathsf{F}_1 ]  \langle \neg\mathsf{F}_1\rangle{\tt t}.$$
In fact, $\mathsf{G}$ is not satisfied by $\mathsf{P}_0$, written  $\mathsf{P}_0 \not\entails \mathsf{G}$, because, along the unique possible path,  the labels of the first two transitions satisfy $\neg\mathsf{F}_1$ but $\mathsf{P}_2$ cannot perform any transition whose label satisfies $\neg\mathsf{F}_1$.

Differently, $\mathsf{P}_0' \entails \mathsf{G}$. In fact, $\mathsf{P}_0'$ can move to $\mathsf{P}_1'$ with a transition whose label satisfies $\neg \mathsf{F}_1$, then $\mathsf{P}_1'$ has a unique transition to $\mathsf{P}_2'$ whose label satisfies $\neg \mathsf{F}_1$ and finally the target state $\mathsf{P}_2'$ can perform a transition whose label satisfies $\neg\mathsf{F}_1$.
\end{example}

\section{Implementation}
\label{sec:impl}

In Falaschi and Palma~\cite{FP2020} we have presented some preliminary work on how to implement RS formalism in a logic programming language (Prolog). Our implementation did not aim to be highly performing. We aimed to 
obtain a rapid prototyping tool for implementing extensions of Reaction Systems. 
Our initial
prototype allowed to perform finite computations on RSs, in the form of interactive processes. Here we have extended the implementation by including the more general notion of contexts, the labels and keeping track of them building corresponding LTSs. Then we have added the predicates for formulating  expressions of our assertion language that acts on the transition labels. On the basis of this assertion language we have implemented a slightly modified version of the Hennessy-Milner logic to make it parametric on the specific assertion specified by the user. 
Our interpreter is available for download\footnote{
\url{https://www3.diism.unisi.it/~falaschi/AssertionsForReactionSystems}}.

For performance reasons and in conformance with the double-arrow transition system, our implementation uses the ({\em InH}) rule in a \emph{deterministic} way by maximising the sets of present inhibitors and lacking reagents in the current computation. This improves the efficiency of the tool.
We have run and checked the examples in this paper, by using our interpreter.
As explained in the  online instructions, the tool can be easily customised by instantiating a few predicates providing,
respectively, the Reaction System specification and a BioHML formula to be verified.


\section{Two extensions}
\label{sec:extension}

Here we present two extensions: a numeric extension that takes into account
the number of times an entity  is used as a reactant in a single transition;
an extension that introduces an operator for letting two RSs be \emph{connected}.

\medskip

\noindent
{\bf Reactant occurrences.}\\
The first idea is to introduce some naive measure for the number of entities that are needed by the reactions.
Now, we assume that
the number  associated to entities in the sets $R$ (reactants) and $P$ (products) are the 
stoichiometric  numbers, 
as specified in the corresponding biochemical equation.
This amounts to use multisets instead of sets (for $R$ and $P$) within the labels.
The set $I$ (of inhibitors)  remains a simple set. 
At the level of notation, we write a multiset as a formal sum
$\bigoplus_{\mathsf{a}\in S} n_{\mathsf{a}}\mathsf{a}$,
where $n_{\mathsf{a}}\in \mathbb{N}$ is the number of occurrences of $\mathsf{a}$.
For simplicity, we write just $\mathsf{a}$ instead of $1\mathsf{a}$ and we omit any term of the form $0\mathsf{a}$.
For example, the multiset $2\mathsf{a}\oplus \mathsf{b}$ has two instances of $\mathsf{a}$ and  one of $\mathsf{b}$.
Overloading the notation we use $\cup$ as multiset union, i.e.
$$(\bigoplus_{\mathsf{a}\in S} n_{\mathsf{a}}\mathsf{a}) \cup (\bigoplus_{\mathsf{a}\in S} m_{\mathsf{a}}\mathsf{a}) = \bigoplus_{\mathsf{a}\in S} (n_{\mathsf{a}}+m_{\mathsf{a}})\mathsf{a}$$
If $R=\bigoplus_{\mathsf{a}\in S} n_{\mathsf{a}}\mathsf{a}$ we let $R(a)=n_{\mathsf{a}}$.

Similarly, we want to use multisets also for the contexts, but in this case
we want the possibility to parameterize the context  w.r.t. the number of entities it provides.
To this purpose, fixed a finite set $X=\{x_1,...,x_n\}$ of variables, 
we introduce some linear expressions of the form $e = \sum_{i=1}^n k_ix_i + h$
with coefficients $k_i,h\in \mathbb{N}$, such that  a context $C$ associates to each entity $\mathsf{a}$ a linear expression $e_{\mathsf{a}}$ and not just a number.
Thus we write a context $C$ as a formal sum
$C= \bigoplus_{\mathsf{a}\in S} e_{\mathsf{a}}\mathsf{a}$.
A multiset is just a particular case of the above expression where all variable coefficients are $0$.
For example, we can let $C = (x+y)\mathsf{a} \oplus (x+1)\mathsf{b}$.
The union of contexts is then defined as follows
$$
\bigoplus_{\mathsf{a}\in S} e^1_{\mathsf{a}}\mathsf{a} \cup 
\bigoplus_{\mathsf{a}\in S} e^2_{\mathsf{a}}\mathsf{a}
=
\bigoplus_{\mathsf{a}\in S} (e^1_{\mathsf{a}}+e^2_{\mathsf{a}})\mathsf{a} 
$$

We assume that variables in $X$ can only range over positive values, so that if $e_{\mathsf{a}} \neq 0$ then $\mathsf{a}$ is present in $\bigoplus_{\mathsf{a}\in S} e_{\mathsf{a}}\mathsf{a}$.

In the SOS rules we need to use the requirements $(W\cup R) \cap I = \emptyset$ and $R\subseteq W$. They are intended to be satisfied at the qualitative level, not necessarily at the quantitative one. Correspondingly, the disjointness condition $(W\cup R) \cap I = \emptyset$
 is satisfied when $\forall \mathsf{a}\in I.~ (W\cup R)(\mathsf{a})=0$, and the inclusion condition
$R\subseteq W$ is satisfied when
$\forall \mathsf{a}\in S.~ R(\mathsf{a})\neq 0\Rightarrow W(\mathsf{a})\neq 0$.
Our new transition labels  differ from the ones in Figure~\ref{fig:LTSforRS} just because $R$, $P$, and $W$ are now multisets.
We keep the same SOS rules as before.

The advantage is that to each transition $\mathsf{P}\xrightarrow{\obs{W}{R,I,P}}\mathsf{P}'$ we can now assign a system of linear inequalities:
$\forall \mathsf{a}\in S.~ R(\mathsf{a}) \leq W(\mathsf{a})$, where $R(\mathsf{a})\in \mathbb{N}$ and $W(\mathsf{a})$ is an expression.
%
The aim is to estimate, with no computational effort, the \emph{relative quantities} of biological material which should be provided to the system to reach a  desired  configuration.
This could be helpful during the setting phase of an \emph{in vitro} experiment to avoid over-use of biological material, given its high cost.
Please note that the qualitative nature of RS is unchanged, we only add some extra information that we elaborate by manipulating transition labels, only.
Here we give an intuition with a short example.

\begin{example}
Let us consider the chemical reactions  in Azimi et al.~\cite{ABP14}, Table~3, in particular reactions $(i)$ and $(vii)$; we
will use their formalization in the syntax of RS, by keeping the stoichiometric numbers:
\[
a_1\defeq (\{(\mathsf{hsf},3)\}, \{\mathsf{d_I}\}, \{\mathsf{hsf}_3\}) \qquad 
a_2 \defeq (\{\mathsf{hsp}, \mathsf{hsf_3}\}, \{\mathsf{d_I}\}, \{\mathsf{hsp\mathrm{:}hsf}, (\mathsf{hsf},2)\})
\]
Reaction $a_1$ requires three copies of the entity $\mathsf{hsf}$, while $a_2$ produces two copies of $\mathsf{hsf}$.
We assume that the context initially provides the  set $C\defeq x\mathsf{hsf}\oplus\mathsf{hsp}\oplus \mathsf{hsf_3}$ and then it provides the empty set, i.e. it is defined as $\mathsf{K} \defeq C.\emptyset.\nil$.
The resulting system can only execute two transitions: in the first transition  both reactions $a_1$ and $a_2$ are applied, in the second transition only reaction $a_1$ is applied:
\[
[\mathsf{K} |a_1| a_2] \xrightarrow{\obs{C}{R,I,P}} [P|\emptyset.\nil |a_1|a_2] \xrightarrow{\obs{P}{R',I',P'}} [P'|\nil|a_1|a_2]
\]
\[
\begin{array}{llllll}
\mbox{where}\quad& R =3\mathsf{hsf}\oplus\mathsf{hsp}\oplus\mathsf{hsf_3}& \quad &
I=\{\mathsf{d_I}\} & \quad &
P=\mathsf{hsf_3}\oplus\mathsf{hsp\mathrm{:}hsf}\oplus 2\mathsf{hsf}\\
& R' = 3\mathsf{hsf}& &
I'=\{\mathsf{hsp},\mathsf{d_I}\}& &
P'=\mathsf{hsf_3}
\end{array}
\]
\end{example}
Now, from the first transition we extract the requirement $R(\mathsf{hsf}) = 3 \leq C(\mathsf{hsf}) = x$, while from the second transition we get $R'(\mathsf{hsf}) = 3 \leq P(\mathsf{hsf}) = 2$. If we would wanted  a quantitative estimate of need of  entity $\mathsf{hsf}$, this comparison would reveal
 that the production of $\mathsf{hsf}$ is not sufficient to trigger the second reaction.


\medskip
\noindent
{\bf The connector operator.}\\
In Bodei et al~\cite{BBF19} and Brodo et al.~\cite{TCS:RSLink} we have presented the encoding of RS into the {\tt link}-calculus and we have already discussed how to connect two encoded RS such that some of the entities produced by one RS 
are provided to the second one, similarly to what has been done in Bottoni et al.~\cite{BLR2020}.
To this aim we introduce an operator, that we call ``connector", written as $\mathsf{P}_1 \conn{L} \mathsf{P}_2$, 
meaning that when the RS process $\mathsf{P}_1$ produces entities in the set $L$, these entities are available, at the next step, as reactants to the continuations of both RS processes.
As a special case, when $L=\emptyset$, there cannot be any exchange of entities and $\mathsf{P}_1$ and $\mathsf{P}_2$ run in parallel, but in isolation. We denote this composition by $\mathsf{P}_1 \parallel \mathsf{P}_2$.

\begin{figure}[t]
$$
\infer[\scriptsize(\textit{Lnk})]
{\mathsf{P}_1\conn{L}\mathsf{P}_2 \xrightarrow{\obs{W_1\cup W_2}{R_1\cup R_2,I_1 \cup I_2,P_1\cup P_2}} \mathsf{P} \conn{L} [\mathsf{M}|(L\cap P_1)]}
{\mathsf{P}_1 \xrightarrow{\obs{W_1}{R_1,I_1,P_1}} \mathsf{P}
& \mathsf{P}_2 \xrightarrow{\obs{W_2}{R_2,I_2,P_2}} [\mathsf{M}]}
$$
\caption{SOS semantics rule for the connector operator}
\label{fig:newLTSforRS}
\end{figure}

\section{Conclusion and future work}
\label{sec:conc}

We have presented an SOS semantics for the Reaction Systems that generates a labelled transition system.
We have revised RSs as processes, formulating a set of ad-hoc inference rules.
In a way we have a  flexible framework that allows one to add new operators in a natural way.
It is important to note that the transition labels  play an interesting role, not only because they reflect
the important aspect of  the computations, but also because they can add expressivity at the computation
allowing for additional analysis, as we did in Section~\ref{sec:biosimulation}.
In Section~\ref{sec:impl} we have briefly described a preliminary interpreter in logic programming which implements the verification of BioHML formulas on computations of RS processes with nondeterministic contexts in our framework.

As future work we plan to apply our technique to define SOS semantics for other synchronous rewrite-rule systems (where all the rules are applied synchronously) to define a uniform computational framework. We also plan to improve our implementation including its functionalities, interface and usability.

\medskip

\noindent
{\small {\bf Acknowledgments}
We thank the anonymous reviewers for their detailed and very useful 
criticisms and recommendations that helped us to improve our paper.}
\bibliography{biblio_CNA}
\bibliographystyle{abbrv}

\end{document}